\documentclass{article}
\usepackage{amsmath,epsfig}
\usepackage{authblk}
\usepackage[shortlabels]{enumitem}

\title{Multispectral to Hyperspectral using Pretrained Foundational model}

\author[1]{Ruben Gonzalez*}
\author[1]{Conrad M Albrecht}
\author[1]{Nassim Ait Ali Braham}
\affil[1]{Remote Sensing Technology Institute, German Aerospace Center (DLR), Germany}

\author[2]{Devyani Lambhate*}
\author[2]{Joao Lucas de Sousa Almeida}
\author[2]{Paolo Fraccaro}
\author[2]{Benedikt Blumenstiel}
\author[2]{Thomas Brunschwiler}
\author[2]{Ranjini Bangalore}
\affil[2]{IBM Research Labs, India, U.K., Zurich, Brazil}

\begin{document}

\maketitle

\begin{abstract}
Hyperspectral imaging provides detailed spectral information, offering significant potential for monitoring greenhouse gases like CH\textsubscript{4} and NO\textsubscript{2}. However, its application is constrained by limited spatial coverage and infrequent revisit times. In contrast, multispectral imaging delivers broader spatial and temporal coverage but lacks the spectral granularity required for precise GHG detection. To address these challenges, this study proposes Spectral and Spatial-Spectral transformer models that reconstructs hyperspectral data from multispectral inputs. The models in this paper are pretrained on EnMAP and EMIT datasets and fine-tuned on spatio-temporally aligned (Sentinel-2, EnMAP) and (HLS-S30, EMIT) image pairs respectively. Our model has the potential to enhance atmospheric monitoring by combining the strengths of hyperspectral and multispectral imaging systems.

\end{abstract}
%
%
\section{Introduction}
\label{sec:intro}
Satellite images are being used to create detailed maps of Earth’s surface. These maps can be used for a variety of purposes, including navigation, urban planning, and environmental management. Recently there has been a surge in the launch of Hyperspectral satellites. Unlike multispectral satellites that capture a few broad bands, hyperspectral satellites capture hundreds of narrow bands, providing a much richer spectral fingerprint. Because of its ability to capture finer spectral information, it can be used for more precise applications like mineral mapping, greenhouse gases (GHGs) mapping, precision agriculture, surveillance, water resource management, and many other tasks. Hyperspectral imaging has revolutionized many fields by enabling a detailed analysis of different matters and molecules based on spectral signatures. However, the revisit time of most of the hyperspectral satellites are much higher, which can be a limitation for time sensitive observations. 

 In contrast multispectral satellites have lower revisit time and broader spatial coverage. In this paper, we propose to generate Hyperspectral data from Multispectral data, for the periods when hyperspectral data is unavailable using foundational models.

 Recently large-scale Image foundational models like Vision Transformer(ViT) \cite{vit}, Stable Diffusion \cite{rombach2021highresolution}, and DALL·E \cite{ramesh2021zeroshottexttoimagegeneration} have revolutionized image generation, image  classification and segmentation tasks. We have formulated the problem of generating hyperspectral data from multispectral data as a precise data generation case, where we want to generate all hyperspectral bands given the multispectral bands for a satellite Image using a ViT model. The ViT models focuses on capturing spatial relationships but lack the ability to capture spectral relationships. We therefore propose two modified version of vision transformers, specially adapted to capture the spectral and spatial-spectral relationships in the hyperspectral data.

We have performed two set of experiments to generate hyperspectral data from the multispectral data.
\begin{enumerate}

    \item  Multispectral data from Harmonized Landsat Sentinel-2 (HLS-S30) \cite{CLAVERIE2018145} is used to reconstruct the corresponding Earth Surface Mineral Dust Source Investigation (EMIT) \cite{emit} hyperspectral bands using a ViT-based model pre-trained on EMIT data. 
    \item Multispectral data from Sentinel-2 (S2) is used to reconstruct the corresponding Environmental Mapping and Analysis Program (EnMAP) \cite{enmap} using a ViT-based model pre-trained on EnMAP data.

\end{enumerate}

\begin{figure*}[t!]
    \centering
    \includegraphics[width=\textwidth]{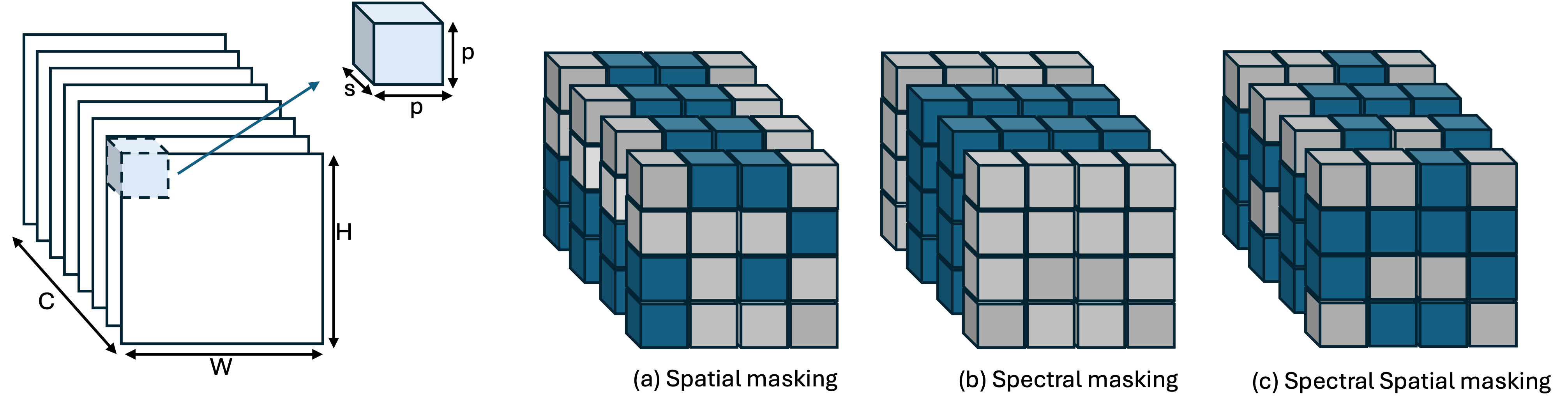} 
    \caption{Masking strategies}
    \label{fig:fullpage}
\end{figure*}
\section{Related work}
Spectral reconstruction bridges the gap between hyperspectral and multispectral imaging by enhancing spectral granularity. Early methods, such as dictionary-based approaches \cite{arad-2016} and Gaussian processes \cite{akhtar-2018}, relied on handcrafted priors but were limited in capturing non-linear spectral-spatial relationships. With the rise of deep learning, convolutional neural networks (CNNs) \cite{galliani-2017, Xiong_2017_ICCV} and attention-based transformers \cite{cai-2021} have emerged as effective tools, leveraging data-driven learning to improve reconstruction accuracy. Recent approaches, such as hybrid attention networks \cite{zheng-2021}, emphasize combining spatial and spectral attention for enhanced reconstruction quality. 

Self-supervised learning (SSL) frameworks, like masked autoencoders (MAE) \cite{mae}, naturally align with spectral reconstruction by leveraging large-scale unlabeled datasets to predict missing spectral information \cite{hackstein2024}. In hyperspectral remote sensing, studies such as \cite{linus2023SST} have demonstrated the effectiveness of self-supervised masked image reconstruction for adapting transformers to the unique characteristics of hyperspectral data. By integrating spatial-spectral self-attention, spectral positional embeddings, and blockwise patch embeddings, these models achieve significant accuracy improvements, particularly in label-scarce scenarios. Building on these insights, our work further explores spectral reconstruction with self-supervised transformers tailored for hyperspectral imagery.

\section{Data}

\subsection{Pretraining datastes}

\textbf{EMIT} uses an advanced imaging spectrometer instrument installed on the International Space Station that measures a spectrum for every point in the image. The EMIT data reflects the wavelengths from visible to short infrared wavelengths consisting 285 discrete bands, we have used 240 bands for our experiments with excluding the bands corresponding to water vapours and the first four bands. We use the Level 2A EMIT \cite{thompson2020emit} product that contains the Surface reflectance derived by screening clouds and correction for atmospheric effects. The EMIT dataset used for Pretraining is split into 73,853 train, 4102 validation and 4102 test samples.  \\

\textbf{{EnMAP}} is derived from two primary sources. HySpecNet-11k provides 11,483 hyperspectral images extracted from EnMAP tiles, covering wavelengths from 420 to 2,450 nm and comprising 202 usable spectral bands after atmospheric correction \cite{fuchs2023hyspecnet11klargescalehyperspectraldataset}. This dataset offers a manageable yet diverse subset of hyperspectral data for model training and evaluation. Complementing this, the SpectralEarth dataset comprises 538,974 hyperspectral images derived from EnMAP acquisitions between 2022 and 2024. With similar spatial and spectral characteristics to HySpecNet-11k, SpectralEarth provides extensive geographic and temporal coverage, making it well-suited for large-scale pre-training \cite{braham2024spectralearthtraininghyperspectralfoundation}.

\subsubsection{Finetuning Datasets}
\textbf{HLS-S30 and EMIT} images from a harmonized surface reflectance product called HLS-2 (version 2.0) data are downloaded from NASA's land processes distributed active archive center cumulus cloud as optimized GeoTIFFs. The Sentinel S30 data has been pre-processed to generate and select 128 $\times$ 128 pixel$^{2}$ cloud-free patches. The data is processed to spatio-temporally align the EMIT and Sentinel-S30 pairs. This dataset is divided into 20691 train, 3205 validation, and 2190 test patches. Sentinel-S30 cloud-free patches with 13 channels were utilized to generate the corresponding 240 hyperspectral bands.

\textbf{S2 and EnMAP}
The multispectral imagery is sourced from Sentinel-2, aligned spatially and temporally with the hyperspectral data of the HySpecNet-11k dataset. Level-2A-processed Sentinel-2 images were obtained via Google Earth Engine and matched to EnMAP acquisitions within a 3-day window to ensure comparable surface reflectance values. After cloud filtering, this mapping resulted in an 85.32\% alignment between EnMAP patches and their corresponding Sentinel-2 images. 

\label{sec:format}

\label{sec:pagestyle}

\begin{table}
  \centering
  \begin{tabular}{ |c|c|c|c|c|}

    \hline
     \multicolumn{1}{|c|}{Masking}  & \multicolumn{3}{|c|}{MSE} &{SSIM} \\ \cline{1-5} 
    Train  &{Total} &{Masked}
    &{Unmasked}
    & \\
    \hline
    Spec  &7.11 E-4 & 9.17 E-4 & 9.25 E-5  &  0.7372 \\
    Spa-Spec  &\textbf{4.27 E-4} &\textbf{5.6 E-4} & \textbf{2 E-5}  &\textbf{0.8571} \\
    
    \hline
  \end{tabular}
   \caption{ A comparison of Spectral (spec) and Spectral-Spectral (Spa-Spec) masking strategy with EMIT dataset. In these experiments, random 75\% band groups are masked (similar to spectral masking) to evaluate the test set.}
  \label{tab:1}
\end{table}

\section{Pretraining Hyperspectral Foundational Model}
\subsection{Model Architecture}
We used a very small version of the ViT model based on the MAE approach, a successful self-supervised learning method widely used and extended for different data types, including videos\cite{videomae} and multispectral images \cite{satmae}. The MAE reconstructs masked images using an asymmetric encoder-decoder architecture. The input image is divided into non-overlapping patches of the same size, and a subset of the patches is randomly masked. The encoder receives only the unmasked patches generating their latent representation. The decoder then receives the latent and masked tokens to perform the image reconstruction task \cite{mae}.

We have 4 encoder and 2 decoder blocks in our very small ViT model. The number of heads are (8 and 8) and the embedding dimensions are (768 and 512), respectively, for (the encoder and the decoder). The images are $128 \times 128\times$ 240 (H $\times$ W $\times$ C) dimensional. We have trained the model with a masking ratio (r) of 75\%. Unlike the loss function used in ViT, which measures the loss over just masked pixels, we use a holistic loss that measures the loss over the reconstruction of both masked and unmasked pixels. We have seen better convergence using the holistic loss function as compared to the loss over masked pixels.

\subsection{Masking strategies}

The ViT is pre-trained using a masking approach. We have explored two types of masking strategies in this paper: spectral and spatial-spectral. The input hyperspectral images has 3 dimensions, which are two spatial (H, W) dimensions and one spectral dimension (C), where C is the number of channels in the hyperspectral data. The spatial masking, where the input (H, W, C) is divided into patches of dimension (p, p, C) has proved to be very beneficial for several vision-based models as well as some remote sensing models. The problem with spatial masking is that it does not inherently capture spectral relationships. We therefore propose spectral and spectral-spatial masking for band reconstruction.  


\subsubsection{Spectral Masking}
 In Spectral masking strategy, the input (H, W, C) is divided into (p, p, s) patches with C/s band groups. Out of these C/s band groups, we randomly select r\% of the band groups to be masked, where r is the masking ratio. If a band group is selected to be masked then all the spatial patches (p $\times$ p), corresponding to the selected band groups are masked. We have used band groups, instead of single bands here to take care of the increase in computational complexity if single bands were used. 

\subsubsection{Spatial-Spectral Masking}
 Similar to Spectral masking the input (H, W, C) is divided into (p, p, s) patches. And each of the patch of size (p $\times$ p $\times$ s) is either selected to be masked or unmasked randomly based on the masking ratio (r). The pictorial representation of all three masking strategies (spatial, spectral and spatial-spectral) are presented in the Figure \ref{fig:fullpage}.

\section{Experiments and Results}

\subsection{Metrics}

\textbf{MSE}: We have reported the Masked MSE (calculated only over masked pixels), Unmasked MSE (calculated only over unmasked pixels) and the total MSE. 
 \\ \\
\textbf{Structural Similarity Score (SSIM)} \cite{wang2004} : SSIM is a metric for evaluating the quality of digital images\[
\text{SSIM}(x, y) = \frac{(2\mu_x \mu_y + C_1)(2\sigma_{xy} + C_2)}{(\mu_x^2 + \mu_y^2 + C_1)(\sigma_x^2 + \sigma_y^2 + C_2)}
\]
Where, $x$ and $y$ are the two images being compared,  $\mu_x$ and $\mu_y$ are the mean intensities, $\sigma_x^2$ and $\sigma_y^2$ are the variances, $C_1$ and $C_2$ are the small constants to stabilize the division, for $x$ and $y$ respectively and $\sigma_{xy}$ is the covariance of $x$ and $y$.

\subsection{A comparison of spectral and spatial-spectral pre-trained models}
The spatial-spectral and spectral models are pre-trained on the EMIT dataset to get a comparison between the two masking techniques. The results are reported in Table \ref{tab:1}. We notice that the model with spectral-spatial masking outperforms the model with spectral masking.

\begin{table*}[h]
  \centering
  
  \begin{tabular}{ |c|c|c|c|c|c|c|}
    
    \hline
    Dataset & Frozen/FT & Masking & \multicolumn{3}{c|}{MSE} &{SSIM} \\ \cline{4-6} 
    
    & && {Total} &{Masked}
    &{Unmasked}
    & \\
    \hline
    HLS-S30-EMIT & Frozen & Spec& 1.26 E-2 & 1.3 E-2 & 1.15 E-2 & 0.5366\\
     HLS-S30-EMIT & FT & Spec& {1.12 E-2} & \textbf{{1.14 E-2}} & { 1.00 E-2} & { 0.58} \\
     HLS-S30-EMIT & Frozen &Spa-Spec &\textbf{1.09 E-2} & \textbf{1.14 E-2} &\textbf{9.79E-3} &0.5908 \\
     HLS-S30-EMIT & FT & Spa-Spec & 1.2 E-2 & 1.23 E-2 & 1.20 E-2 &  \textbf{0.7508} \\
     \hline
      Dataset & Frozen/FT & Masking & \multicolumn{3}{c|}{Total MSE} &{SSIM} \\ \cline{4-6} 
      \hline
     S2-EnMAP &Frozen &Spec & \multicolumn{3}{c|}{9.96 E-3 } &{0.582} \\
     S2-EnMAP &Finetuned &Spec & \multicolumn{3}{c|}{\textbf{3.89 E-3}} &{\textbf{0.767}} \\

    \hline
  \end{tabular}
  \caption{Frozen and Finetuned model's performance for HLS-S30 and EMIT dataset for Spectral (Spec) and Spatial-Spectral(Spa-Spec) masking}
  \label{tab:2}
\end{table*}

\subsection{Reconstruction of Hyperspectral from Multispectral}

 After getting sufficient confidence in generating randomly masked EMIT bands, we tried to test the models for Multispectral to Hyperspectral conversion. 

For this, we have aligned the Sentinel bands with the EMIT and EnMAP bands based on the frequencies (if there is more than 60\% overlap between the frequencies, then we say there is a match). Wherever the match is found, bands are fixed to be the unmasked bands and the rest are the masked bands.  

\begin{figure}[!h]
    \centering
    \includegraphics[width=\columnwidth]{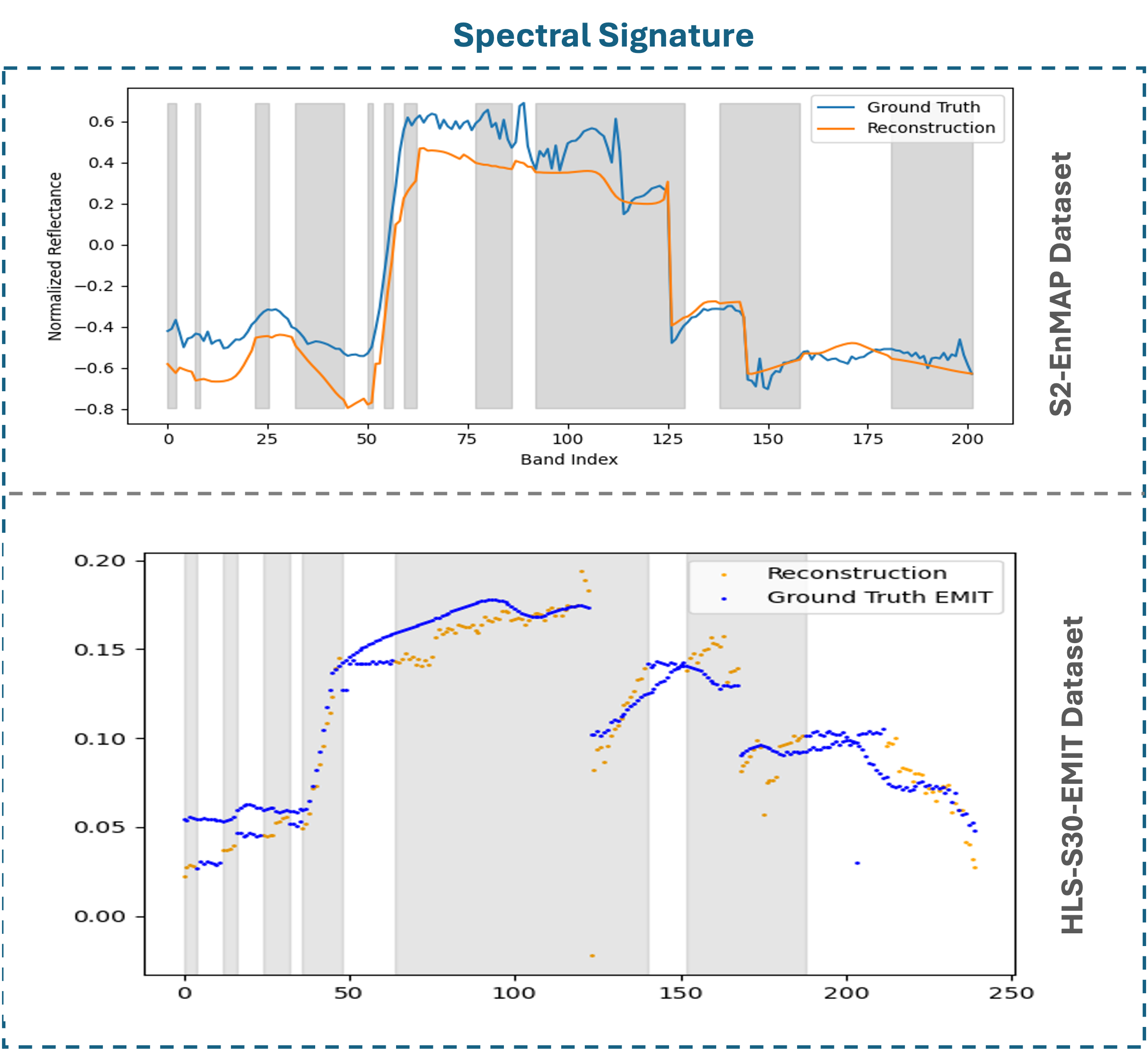}
   
   \caption{Spectral reconstruction for selected pixels for the two datasets. The masked bands are represented by Grey region and the unmasked bands are represented by White region.}
   \label{fig:spec_sign}
\end{figure}

\begin{figure}[!h]
    \centering
    \includegraphics[width=\columnwidth]{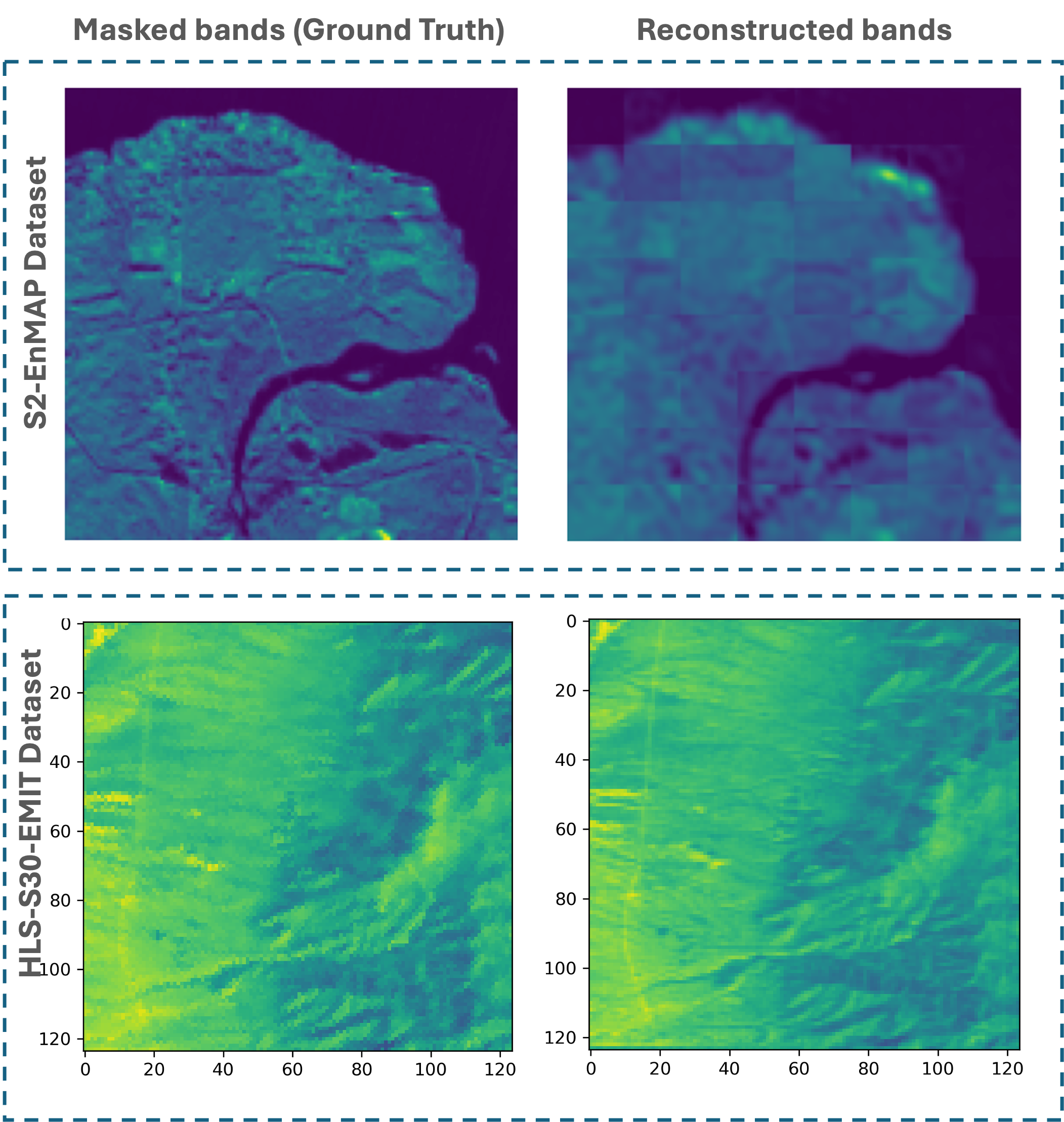}
\caption{Original and reconstructed masked bands for the two datasets}    \label{fig:reconstruction}

\end{figure}

Table \ref{tab:2} summarizes the experiments for band reconstruction for the two data sets with and without fine-tuning. We have selected Frozen Spa-Spec model for the HLS-S30-EMIT dataset and Finetuned Spec model for the S2-EnMAP dataset to generate the plots in Figure \ref{fig:spec_sign} and \ref{fig:reconstruction} as they performed best with-respect-to the MSE metric. The frozen version have never seen any train samples of Sentinel data but still perform considerably well both in terms of MSE metric and reconstruction quality which can be seen from Figure \ref{fig:reconstruction}-HLS-S30-EMIT plot. In all the spectral model versions, the results are further improved by fine-tuning. Whereas, with HLS-S30-EMIT dataset, we have seen an improvement only in the SSIM metric and not MSE for spatial-spectral version. We have not included the results for spectral-spatial generation for the S2-EnMAP dataset as it is already covered in \cite{linus2023SST}.




\section{Conclusion and Future Work}
This study introduced the spectral and spatial-spectral transformer models for reconstructing hyperspectral data from limited spectral inputs, demonstrating accuracy and practical utility. The model effectively predicted missing spectral bands in masked hyperspectral data and enhanced spectral details when applied to multispectral inputs. 

Our future plans include using the pre-trained models described in this paper for several downstream tasks like methane detection, CO2 detection, Crop identification, Mineral classification, etc. We plan to further generalize the masking approach by including a temporal dimension in the masking strategy. We also plan to work on multimodal models, which can process data from multiple hyperspectral/multispectral datasets.

\label{sec:ref}

\bibliographystyle{IEEEbib}
\bibliography{strings,refs}

\end{document}